\documentclass[useAMS,usenatbib,usegraphicx]{mn2e}

\title[ Dense Clouds in a Barred Galaxy]
  { Dense Cloud Formation and   Star Formation in a Barred Galaxy}
\author[M. Nimori et al.]
  {M.~Nimori$^1$, A.~Habe,$^1$, K. ~Sorai$^1$,Y.~Watanabe$^2$,  A.~Hirota$^3$, 
and    D. Namekata$^4$
\\
  $^1$Department of Physics, Graduate School of Science,Hokkaido University, Kita 10 Nishi 8, Kita-ku, Sapporo, 060-0810, Japan\\
  $^2$Department of Physics,  Graduate School of Science, The University of Tokyo, 7-3-1 Hongo, Bunkyo-ku, Tokyo 113-0033, Japan\\
  $^3$ Nobeyama Radio Observatory, 462-2 Nobeyama, Minamimaki, Minamisaku, Nagano 384-1305, Japan\\
  $^4$Center for Computational Science, University of Tsukuba, 1-1-1, Tennodai, Tsukuba, Ibaraki 305-8577, Japan\\}
\date{Released 2012 Xxxxx XX}
\pagerange{\pageref{firstpage}--\pageref{lastpage}} \pubyear{2012}

\def\LaTeX{L\kern-.36em\raise.3ex\hbox{a}\kern-.15em
    T\kern-.1667em\lower.7ex\hbox{E}\kern-.125emX}

\begin{document}

\label{firstpage}

\maketitle

\begin{abstract}

We investigate the properties of massive, dense clouds formed in
a barred galaxy and their possible relation to star formation,  performing
a two-dimensional hydrodynamical simulation with the gravitational
potential obtained from the 2Mass data from the barred spiral galaxy, M83. 
Since the environment for cloud formation and evolution in the bar
region is expected to be different from that in the spiral arm region, barred
galaxies are a good target to study the environmental effects on cloud
formation and the subsequent star formation. Our simulation uses for an initial 80 Myr an
isothermal flow of non-self gravitating gas in the barred potential, then including radiative cooling, heating and
self-gravitation of the gas for the next 40 Myr, during which dense
clumps are formed. We identify many cold, dense gas clumps for which
the mass is more than $10^4M_{\odot}$ (a value corresponding to the
molecular clouds) and study the physical properties of these
clumps. The relation of the velocity dispersion of the identified
clump's internal motion with the clump size is similar to that
observed in the molecular clouds of our Galaxy. We find that the
virial parameters for clumps in the bar region are larger than that in
the spiral arm region. From our numerical results, we estimate star formation
in the bar and spiral arm regions by applying the simple model of \citet{km}.
The mean relation between star formation rate and gas surface density
agrees well with the observed Kennicutt-Schmidt relation. The SFE in the bar region is $\sim $ 60 \% of  the
spiral arm region. This trend  is consistent with observations of barred galaxies.
\end{abstract}


\section{Introduction}

Star formation (SF)  is one of  the key processes governing the evolution
of galaxies. Many observations of nearby disk galaxies indicate an
empirical relation between gas surface density  ($\Sigma_{gas}$) and
star formation rate surface density  ($\Sigma_{SFR}$). This relation
is called the Kennicutt-Schmidt (K-S) relation (\citet{sch},
\citet{kn}). \citet{blwb} show that $\Sigma_{SFR}$ is well correlated with the
hydrogen molecular surface density ($\Sigma_{H_2}$) for local spiral galaxies.

Many  papers have been devoted to understanding the physical reason behind the K-S
relation (e.g.  \citet{km}). However, it remains poorly
understood. Since massive stars are  mainly observed in the spiral
arms of disk galaxies, it has been proposed that star formation is
regulated by galactic shocks driven by spiral density waves
(e.g. \citet{bt})  and by the increase of cloud-cloud collisions  (\citet{tt},  \citet{ts11}). 
\citet{km} propose a star formation model in which star formation
efficiency depends on the turbulent properties of the molecular clouds and demonstrate that the Kennicutt-Schmidt relation can be explained by their model.
Turbulent internal motion in clouds can be excited by cloud-cloud
collisions via the conversion of the orbital energy of the clouds into internal motion energy (\citet{bdrp},  \citet{dbp},  \citet{tt} and  \citet{ts11}).

Barred galaxies show different star formation activity in the bar
regions and in the spiral arm regions, even when gas surface density
in both regions is comparable (e.g.,  \citet{drsr},  \citet{srvt},
\citet{mktk}, and  \citet{moks}). Studying the physical reason for
this discrepancy is an important step to understanding the physics
for the K-S relation.

It is well known that the star formation activity is higher in spiral
arm regions than in the bar regions, even if the bar regions are gas
rich \citep{moks}.  We focus on the possibility that this difference
is related to a difference in cloud properties between these regions,
since the cloud environment is thought to be closely related with SF
(\citet{km}). 

Gas flows in the bar regions are elongated  with strong shears and dark
lanes that are evidence for strong shock waves　( e.g. \citet{wh}). It is therefore
natural to expect clouds forming in this environment to have different
properties from elsewhere in the disk.



In this paper, we present a numerical simulation of gas flow at high
resolution in a barred galaxy potential. We include cooling and
heating processes and study the properties of clouds in different
galactic environments within the disk and how this relates to star
formation. For the purpose, we resolve gas clumps as small as
molecular clouds.

As the first step in our study, we perform a two dimensional
simulation with   spacial resolution of 4 pc. The galaxy model is
that of a barred galaxy with a bar similar to M83. M83 is a nearby barred galaxy, type SABc,   and one of the
best targets to study the spacial variation of star formation
efficiency in a  barred galaxy, since it is nearby and has been well
observed in various wave lengths, e.g., atomic gas \citep{hb}, the molecular gas (\citet{lwor}, \citet{sm}, \citet{mktk}), optical emission lines \citep {dpt}, X-ray \citep{sw}.  

From the physical properties of the identified clumps, we estimate the SFE and SFR, using the simple star formation model of turbulent clouds proposed by  \citet{km}.

The plan of this paper is as follows: Section 2 describes our model
and numerical method. In section 3, we show our numerical results. In
section 4, we estimate SFR and SFE in the bar region and the spiral arm
region by using the simple model of  \citet{km}. In section 5, we
present our discussion and conclusion.

\section{MODEL and NUMERICAL METHOD}

\subsection{Model Galaxy}

\begin{figure}
  \includegraphics[width=84mm]{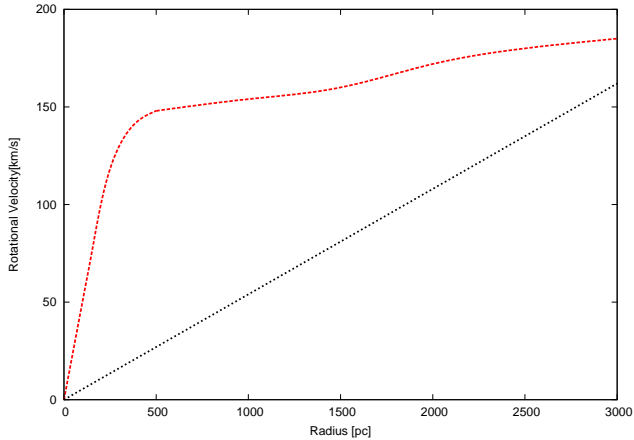}
	 \caption{The circular velocity in the model galaxy and the pattern speed of the bar.}
  \label{f:rotation}
\end{figure}

We use a gravitational potential for a barred galaxy similar to that of
M83.  This model is from the work of \citet{hi} who analyzed the 2Mass K-band image of M83 \citep{jcc}, assuming a constant mass to
light ratio and a distance of 4.5 Mpc to  the galaxy. The model galaxy consists of stellar bulge, stellar bar, 
stellar disk and dark halo components. The stellar bar end  is near $r = 2$ kpc from the galactic center and  loose spiral arms are traced from the  bar end to $r > 4.5$ kpc.  
The circular rotation velocity
of the gravitational potential that is shown in Figure~\ref{f:rotation} roughly agrees  with the observation of
molecular gas in M83 by \citet{lowr}. The observed rotation velocity of molecular gas is 150 km/s at 2 kpc from the center and 180 km/s at 3 kpc from the center  \citep{lowr}. 
In our simulations, we assume that initial gas rotates in the
gravitational potential with a circular velocity that balances the
axial averaged gravity of the barred galaxy. The pattern speed of the
bar potential is $\Omega_{p} = $ 54 km s$^{-1}$ kpc$^{-1}$, in keeping
with the observed global characteristics of the gas distribution in
M83 (\citet{hi}, \citet{hetal}). 

\begin{figure}
  \includegraphics[width=84mm]{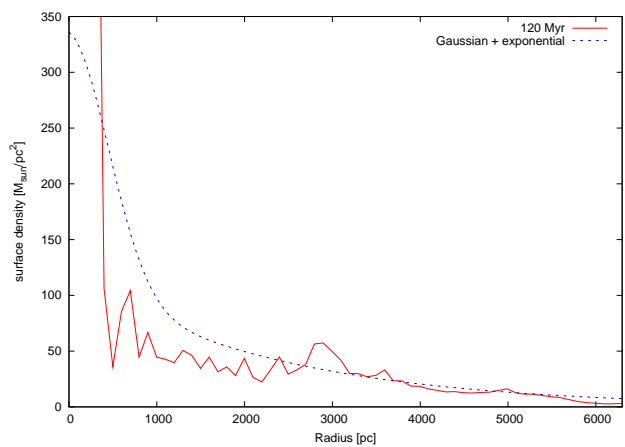}
	 \caption{The azimuthally averaged gas surface density at $t = $ 120 Myr (solid line).
	 	 	Dotted line is the initial gas distribution.}
  \label{f:rsfd120}
\end{figure}

For the  initial radial distribution of gas mass surface density in
the disk, we assume  the Gaussian central component and the
exponential gas disk component given by \citet{lwor}
(Fig.~\ref{f:rsfd120}). The total gas mass within a radius of 6.3 kpc is $3.5 \times 10^9 M_{\odot}$. 

\subsection{Numerical Method}

We simulate two dimensional gas flow in the model galaxy using our
M-AUSMPW$^+$ code \citep{nh}, which is based on an advection upstream splitting  scheme \citep{kkb} with the MLP5 \citep{kka} for calculating the higher oder numerical fluxes.
The size of simulation region is 12.6 kpc $\times$ 12.6 kpc and covers
the whole stellar disk  of M83. The grid size in our simulation is 
3125$^2$, with a cell size of  
4 pc.
We assume as the outer boundary condition of our simulations  that physical quantities of hydrodynamics are continuous.  We have examined this condition by checking that artificial gas motions are not   induced by the outer boundary condition and have found that   outward  gaseous disturbance in a rotating super sonic gas   in the disk  galaxy potential   propagate without reflection at the outer boundary.

For the first 80 Myrs, we calculate gas motion assuming an isothermal,
non self-gravitating gas with $T=10^4$K. At this stage, 
gas density distribution is nearly steady.  
After this time, we take into account
radiative cooling, heating and self-gravity of the gas for the next 40
Myr, which is long enough for the formation of the molecular clouds. 
Self-gravity of the gas is calculated by FFT of which  a detailed description  is given in \citet{nh}.
We do not include star formation processes or any stellar feedback
throughout our simulation, as we are concentrating on the properties
of the clouds formed in the bar galaxy potential. 
We use a cooling function of gas with a solar metallicity given by \citet{sn}.  
For the heating processes, we assume a uniform FUV radiation field and a uniform cosmic ray heating in the energy conservation equation in the numerical hydrodynamics for simplicity. 
We use the far ultra-violet (FUV) heating rate $\Gamma_{\rmn{FUV}} =
1.0 \times 10^{-24} \varepsilon G_0$ \citep{gi}. 
We assume $\varepsilon =0.05$ and $G_0=1.0$ \citep{hb}. 
These values are obtained in the Galaxy. When radiative cooling is
allowed, the gas is allowed to cool to 10 K. For calculation of the cooling rate, gas density is obtained by assuming the thickness of the gas disk is 70 pc. 



\subsection{Clump finding method}

In order to study the properties of clumps in the bar and spiral arm  region, we
identify clumps that are dense and have  a low temperature by the following
method: (A more detailed description  is given in \citet{nh}).
First, cells with a surface density higher than $\Sigma_\rmn{cl}$ and
temperature lower than $T_\rmn{cl}$ are selected as a 'candidate'
clump member.
Next, if a neighbouring cell to  the 'candidate' is also a candidate,
these cells become members of the same group. 
We iterate this procedure for all candidate cells. 
If cell number of a group exceeds $n_\rmn{cl}$, the group is identified as a clump. 

Since the mean surface gas density is less than $60
$M$_{\odot}$/pc$^2$ except for where  $r<500$ pc, as shown in
Figure~\ref{f:rsfd120}, we assume
$\Sigma_\rmn{cl}=60$M$_{\odot}$/pc$^2$ and $n_\rmn{cl}=12$. These  values of $\Sigma_\rmn{cl}$ and $n_\rmn{cl}$ correspond to the  lower limit mass  of clumps of  $m_\rmn{cl}\sim 10^4M_{\odot}$. 

\section{RESULTS}

\subsection{Disk gas evolution}
\begin{figure}
{ 
  \includegraphics[width=85mm]{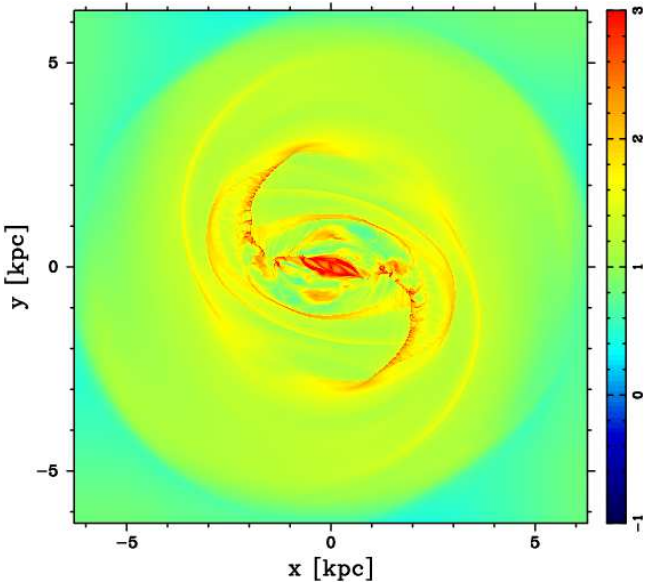}}
	 \caption{The gas surface density map at $t = $ 80 Myr. Until this stage, gas is isothermal. The color bar in the right hand side shows the logarithmic scale of  the surface density of gas in unit of $M_{\odot}$ pc$^{-2}$  }
  \label{f:sfdm80}
\end{figure}

Figure~\ref{f:sfdm80} shows the gas surface density distribution at
the final stage of the  isothermal gas simulation at $t =$ 80 Myr. In this Figure, rotation of  the gas and the bar pattern speed are in the counter-clock direction.  The major axis of the bar is along the x-axis.
Two dominant gas spirals extend from the bar ends and dense gas distributes in the bar region.
We continue the simulation with cooling and heating processes and
self-gravity of the gas, untill $t=$ 120 Myr.

\begin{figure}
  \includegraphics[width=85mm]{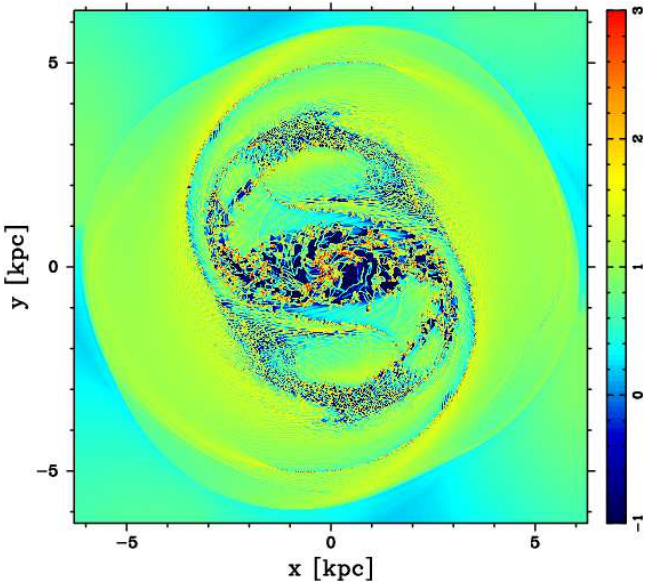}
	 \caption{The same as figure \ref{f:sfdm80}, but for gas surface density map at $t = $ 120 Myr. Our numerical simulation includes radiative cooling  after $t=$ 80Myr.}
  \label{f:sfdm120}
\end{figure}

Figure~\ref{f:sfdm120} shows the gas surface density distribution
at $t =$ 120 Myr.
Many irregular structures of dense gas have formed in the spiral arm  and
bar regions. In Figure~\ref{f:sfdm120}, dense gas extends in the bar region more  than in
Figure~\ref{f:sfdm80}  and there are many  dense gas clumps in the down stream side of the gas flow   in the bar
region  similar to those 
 observed in a barred galaxy. 
 The    gaseous bar extends to  $r=$2kpc  from the center and 
 the loose gaseous spiral arms   begin from the gaseous bar ends and extent to $r=$4kpc.  These features roughly agree with   M83. 
 These dense gas features have a
low temperature ($T < 100$ K) and are similar to the previous studies (\citet{wn01}, \citet{wk}).  
From our numerical result as shown in Figure~\ref{f:sfdm120} we call a rectangular region ($ -2$ kpc $< x <$ 2 kpc  and  $- 1$ kpc$ < y <$1 kpc)  the bar region and a ring region between $r=  2$ kpc and $r=$4 kpc the spiral arm  region.
There are many dense gas cells with $\Sigma > 60M_{\odot}$pc$^{-2}$ and  $T < 100$ K.  Total number of the dense gas cells is 341,854 at $t =$ 120 Myr.

\begin{figure}
  \includegraphics[width=84mm]{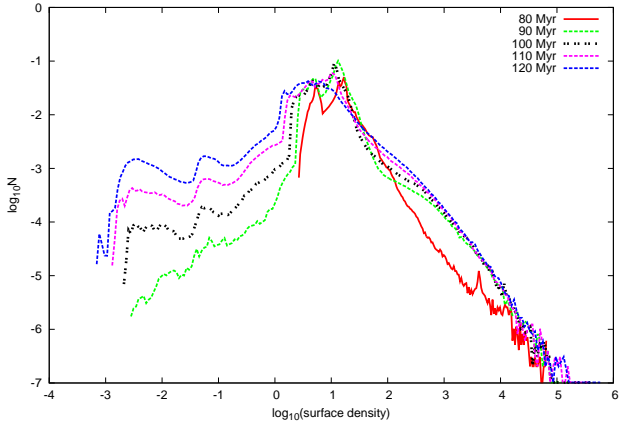}
	 \caption{The PDF of surface gas density in unit of $M_{\odot}/$pc$^{-2}.$ We plot  numerical results at $t=80, 90, 100, 110$, and 120 Myr.}
  \label{f:dpdf}
\end{figure}

We show the probability distribution function (PDF) of surface density
in Figure~\ref{f:dpdf}. This plot excludes the galactic central region
of $r<600 $pc where there is a large concentration of gas. The Figure
shows that the PDF rapidly changes from $t=$80 Myr to 100 Myr and is
almost steady in $t\ge100$ Myr in the range of $\Sigma > $10 $M_{\odot}$/pc$^2$.  The PDF  in the range of    $\Sigma < $1 $M_{\odot}$/pc$^2$ appears in $t> 80$Myr  and is produced  by the FUV heating. The FUV heating hardly affects  dense part of  the PDF in the range of $\Sigma > $10 $M_{\odot}$/pc$^2$.
The PDF in this stage is approximated
by a log-normal form ( \citet{pn},  \citet{wn01}) in the range of $\Sigma > $ 100$M_{\odot}$/pc$^2$. The following analysis focusses
on the simulation results at $t=120$Myr, since the time between $t=80$
and $t=120$ Myr is long enough  to have allowed  more than one free-fall time for a
typical giant molecular cloud (GMC), e.g. $t_{ff}=\sqrt{3\pi/32G\rho}=4.74(n_H/100{\mathrm cm}^{-3})^{-0.5}$Myr, and 
current estimates suggest that  GMCs live between 1 and 2 free-fall times (e.g. \citet{mk07}, and references therin).

We show the radial distribution of the azimuthally averaged gas surface density at $t =$ 120 Myr  in Figure~\ref{f:rsfd120}.
There are gas concentrations in the galactic central region ($r < 500 $pc) and 
near  the bar ends ($r\sim 3$ kpc).  These concentrations are formed by gas dynamical effect by the bar potential (e.g., \citet{at}, \citet{wh}). 
These characteristic features  agree  with the observed radial profile of CO ($J=1-0)$ and CO ($J=3-2$) in M83 \citep{mktk},
except for the large concentration of gas in the central 500 pc region in our numerical result.
In M83, there is a nuclear star burst region at the galactic  center.
 The large concentrated gas within 500 pc in our numerical result  may be consumed by the nuclear star burst and the discrepancy can be small by the burst star formation.

\subsection{Physical properties of the clumps}
We select clumps from our numerical results by the method described in section 2.3.
These clumps are in the mass range of $10^4 <M<10^9 M_{\odot}$.
The most massive clump is in $r < 100$ pc from the galactic
center. The total mass of the clumps is 2.0 $\times 10^9 M_{\odot}$ in the gas disk. The mass of the clumps  in the bar region is 
 1.3$\times 10^9 M_{\odot}$  and  in the spiral arm region is 0.7 $\times 10^9 M_{\odot}$. The total number of clumps in the gas disk is 4122.
 The number of the clumps in the bar region is 495 and in the spiral arm region is 3771.



\begin{figure}
  \includegraphics[width=84mm]{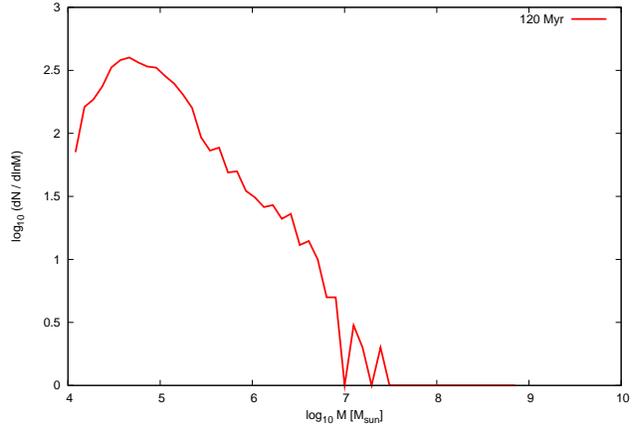}
	 \caption{The clump mass function at $t=$ 120 Myr. }
  \label{f:clmfunc100}
\end{figure}

We show  a mass function of clumps in our numerical result at $t=120$Myr in Figure \ref{f:clmfunc100}. 
In this Figure, we plot the number of clumps with  an equal spacing of $d\log _{10} M=0.05$. 
From Figure \ref{f:clmfunc100}, $dN/dM \propto M^{-1.75}$ in $M> 10^5M_{\odot}$,
since
\begin{equation}
{dN\over dM}={1\over M}{dN\over d\ln M}.
\end{equation}
This power index  agrees well with the observed values $\sim -1.5$ in $M > 10^5M_{\odot}$ in our Galaxy \citep{srby}. We test  different thresholds of surface density  for the clump finding method, and confirm that  the power index of the clump mass function is almost independent from the choice of the threshold value. 

\begin{figure}
  \includegraphics[width=84mm]{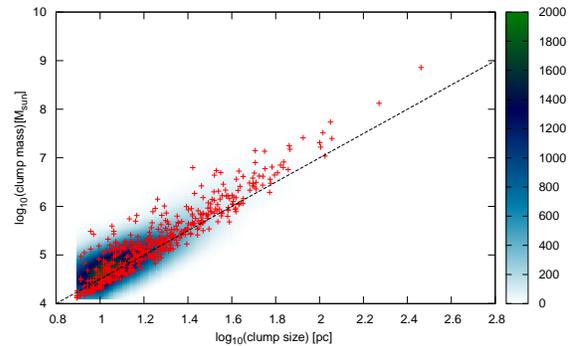}
	 \caption{Clump mass vs.  clump size for the  clumps in our numerical results at $t=$ 120 Myr. 
	 		Plus symbols  show clumps in the bar region and the color contour map   shows  clumps in the spiral arm region. The color  bar shows   number of clumps    in  the  cell     given by  dividing  the plot flame  into     200 $\times$ 200. }
  \label{f:clrclm100}
\end{figure}

We plot clump mass vs. their size in Figure  \ref{f:clrclm100}. 
Plus symbols  show clumps in the bar region.
 We use the color contour map to show clumps in the spiral arm region which
 are too numerous to represent as points. The color bar shows     number of clumps   in  the  cell     given by  dividing  the plot frame  into     200 $\times$ 200.
Since the clumps have an irregular shape, in this Figure,  we estimate clump size $l$ as:
$$
l =(({S\over pc^2})/\pi)^{0.5}pc,
$$ 
where $S$ is  area of the  clump.
From Figure \ref{f:clrclm100}, we find the relation  
\begin{equation}
M_{cl} =10^{4.5} ~( l/10pc)^{2.5}M_{\odot},
\end{equation}
that is shown by a dashed line  in Figure \ref{f:clrclm100}.
Large size clumps of $l >40$ pc are in the bar region and in the galactic central region ($r < 500$ pc).

\begin{figure}
  \includegraphics[width=84mm]{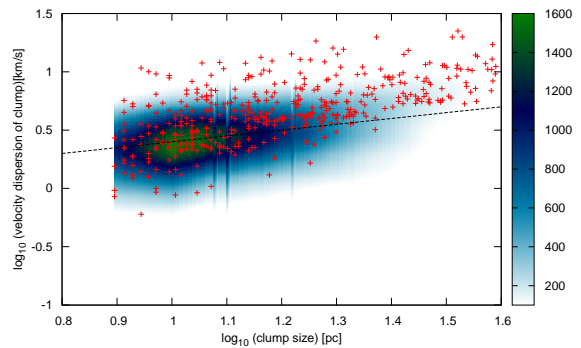}
	 \caption{The same as Figure \ref{f:clrclm100}, but for the distribution of  velocity dispersion of internal motion vs. clump  size  for the  clumps in our numerical results at $t=$ 100 Myr. 
	The dashed line shows the relation given by \citet{srby} 	}
  \label{f:clrvdis100}
\end{figure}

In Figure \ref{f:clrvdis100}, we plot the one dimensional velocity
dispersion, $\sigma _v$, of the internal velocity of the clumps vs. their sizes.
$\sigma _v$  is given by: 
\begin{equation}
\sigma _v =\sqrt{ \frac{ 1}
				 {2 M_\rmn{cl}}\left(\sum_{i,j } \{
					 m_{\rmn{cl}~i, j}  ({\bmath v _{i,j}}  
					- \overline{ {\bmath v_\rmn{cl}}   })^2 \}\right)}
\end{equation}  
from our numerical result, where  $m_{\rmn{cl}~i, j}$ and $\bmath
v _{i,j}$ are the gas mass and the velocity  of gas in the $(i, j)$ cell belonging to  this clump of which mass is $M_\rmn{cl}$, respectively.   $\bmath v_{cl}$ is velocity of the center of mass of  this clump.
Relation of the velocity dispersions and sizes of clumps is approximated as 
\begin{equation}
\sigma_v=0.8~ ( l/1 pc)^{0.5}km/s \label{ml1}
\end{equation}
 in the spiral arm region,  
which is shown by a dashed line in Figure \ref{f:clrvdis100}.
This relation  is similar to  the results of  \citet{l81} and \citet{srby}.
\citet{l81} gives $\sigma_v=1.1~ ( l/1 pc)^{0.38} $ km/s and \citet{srby} gives $\sigma_v=1.0\pm 0.1~ ( l/1 pc)^{0.5\pm0.05} $ km/s.
Figure \ref{f:clrvdis100} clearly show that there are  clumps with  larger velocity dispersions  in the bar region than in the spiral arm region.
 The　large velocity dispersion of clumps in the bar region  may be due to their formation process;  that is, clumps form  from cooled gas in   gas flow   with strong shear motion in the bar region and/or turbulent internal motion in clouds is excited by cloud collisions.

\begin{figure}
  \includegraphics[width=84mm]{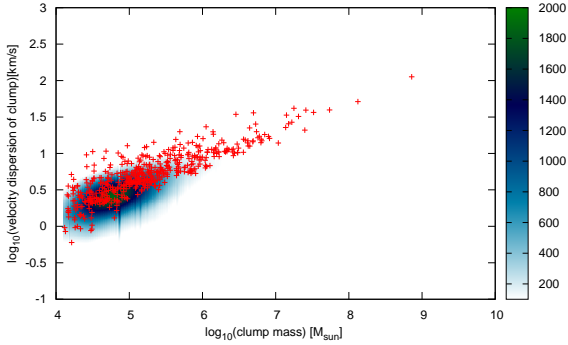}
	 \caption{The same as Figure \ref{f:clrclm100}, 	 but for the distribution of  velocity dispersion of internal motion vs. clump  mass for the clumps in our numerical results at $t=$ 120 Myr. }
  \label{f:clrclvdis_s60n12_100}
\end{figure}

We plot 
internal velocity dispersion  vs. mass of  clumps 
in Figure \ref{f:clrclvdis_s60n12_100}.
The relation between the internal velocity dispersion and the mass of clump
is approximated as 
\begin{equation}
\sigma_v = 2 ~(M_{cl}/10^5M_{\odot})^{0.4-0.5} km/s .
\end{equation}

\begin{figure}
  \includegraphics[width=84mm]{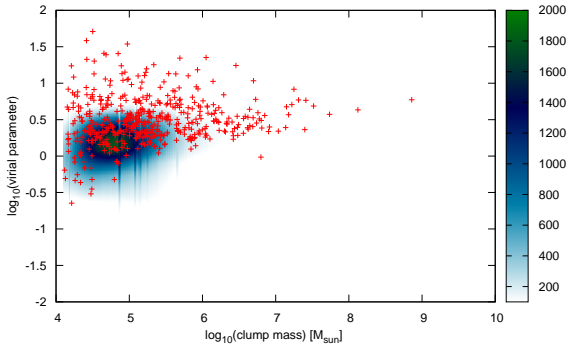}
	 \caption{The same as Figure \ref{f:clrclm100}, but for the distribution of the  virial parameter of clumps with clump mass at $t=$ 120 Myr.}
  \label{f:clralpha100}
\end{figure}

The virial parameter is a useful measure of the gravitational
stability of  the clumps.   
Virial parameter   of a spherical  clump  is given  by  
\begin{equation}
\alpha_{vir} = {{5 \sigma _v^2 l}\over GM_{\rmn{MC}}},
\end{equation}
where   $M_{\rmn{MC}}$ is its mass,   $\sigma _v$ is its one
dimensional velocity dispersion of internal motion,  and $l$ is its
size. 
  We modify $\alpha_{vir}$ as
$$
\alpha_{vir} = {{10 \sigma _v^2 l}\over 3GM_{\rmn{GMC}}},
$$
since our numerical calculation is two-dimensional. 
In Figure \ref{f:clralpha100}, we plot the virial parameter
 $\alpha_{vir}$ vs. mass of clumps.  
 Several clumps in the bar region have larger virial parameters than in the spiral arm region.   Number of clumps with $\alpha_{vir}>1$ is 476  of  the total 495 clumps in the bar and   
 2855  of the total 3771 clumps in the spiral arm region.
  These gravitationally unbound clumps would  be transient, since
  $\alpha_{vir}>1$.
 Such transient clumps are found in  the numerical simulation of giant molecular clouds formation  by  spiral shocks  \citep{dbp}. In \citet{dbp}, they   show that a large part of the formed clouds are  gravitational unbound and transient.　　
 Mass fraction of clumps with $\alpha _{vir} <1$ is 4.28 \% of the total mass of clumps in the disk, 0.33\% of the  total mass of clumps is in the bar region and  3.95\% of the total mass of clumps is in the spiral arm region. Fraction of clumps with $\alpha _{vir} <1$ is much smaller in the bar than that in the  spiral arm region.  
Small fraction of gravitational unstable clumps in the bar region  may be related with   low  star formation efficiency in the bar region. We will discuss this possibility in section 4.

\begin{figure}
  \includegraphics[width=84mm]{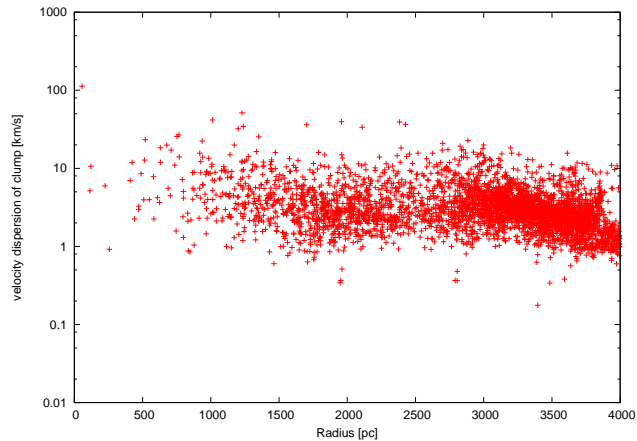}
	 \caption{The radial distribution of   velocity dispersion of clumps in the galaxy at $t=$ 120 Myr.
	 		 }
  \label{f:ralpha_s60n12_100}
\end{figure}

\begin{figure}
  \includegraphics[width=84mm]{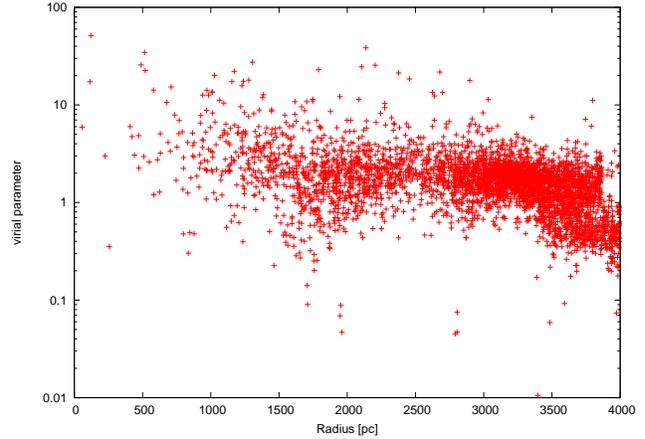}
	 \caption{The  radial distribution of   virial parameter of clumps in the galaxy at $t=$ 120 Myr. 
			}
  \label{f:ralpha100}
\end{figure}

We plot  galactic radial distributions of  velocity dispersion of clumps in Figure \ref{f:ralpha_s60n12_100} and 
   virial parameters of clumps in Figure \ref{f:ralpha100}. 
    There are more clumps with large virial parameters 
    in  $r<2.5$ kpc  than that in $r>2.5$ kpc.
 There are many clumps with $\alpha <1$ 
 in the spiral arm region.   From these results, we expect   many star forming clumps  in the region. 
 In Figure \ref{f:ralpha100}, there are small excess of number of clumps with $\alpha <1$ in $r=1.5-2$kpc. The excess is near the  bar end where gas orbits are expected  to be crowded  \citep{wh}. The orbit crowding will lead to increase cloud collisions near the bar end. The  increase of   cloud collisions can  explain the excess of clumps with
$\alpha <1$, if cloud collisions are inelastic and     reduce internal irregular motion in clumps.


\section{Star formation }
We  estimate the star formation rate from  our numerical results.
Since the cell size of our numerical simulation, 4 pc, is not enough to resolve  the star formation process by hydrodynamical simulation, 
we use  the star formation model  proposed by 
 \citet{km} as sub-grid physics.
 In their model, they assume that star formation   occurs in Jeans unstable  parts of the giant molecular clouds and is regulated by the turbulent motions in the giant molecular clouds.
In their model,   they assume that  turbulent velocities in the   giant molecular clouds is  described by a power   law of   the  size of the turbulent motion,  a probability density function for the density is  a log normal type and star formation occurs  in the Jeans unstable part of  the   giant molecular cloud.   
They obtain 
   a   star formation rate for a giant molecular cloud  as:
 \begin{equation}
SFR=SFR_\rmn{ff} {M_{\rmn{GMC}}\over t_\rmn{ff}}
\end{equation}
where 
\begin{equation}
t_{ff}=\sqrt{{3\pi\over 32 G \rho}}=1.5 \left ({n_H \over 1000 \mathrm{cm}^{-3} }\right )^{-1/2} \rmn{M yr}.
\end{equation}
and 
\begin{equation}
\rmn{SFR_\rmn{ff}}	=
	0.014 \left( \frac{\alpha_\rmn{vir}}{1.3} \right)^{-0.68} \left( \frac{\mathcal{M}}{100} \right)^{-0.32},\label{SFRff}
\end{equation}
where $\mathcal{M} = \sigma_v/c_s$ and $c_\rmn{s}$ is  the sound speed of clouds. 
We estimate  star formation rate in a clump  as
\begin{equation}
\rmn{SFR} = 
	\left( \frac{\rmn{SFR_\rmn{ff}} f_{\rmn{MC}} M_\rmn{cl} }{t_{\rmn{dyn}}}\right) , \label{SFR}
\end{equation}
where  $M_\rmn{cl}$ is the clump mass, 
 $f_{\rmn{MC}}$ is the fraction of molecular gas of it, and 
  $t_\rmn{dyn}$ is   
  given by 
\begin{equation}
t_{\rmn{dyn}} = 	\frac{ \sigma _v}{\pi G \overline{\Sigma}_{\rmn{cl}}}=1.36 \left({\sigma _v \over \rmn{1km~ s}^{-1}}\right)\left({	\overline{\Sigma}_{\rmn{cl}} \over 50M_{\odot}\rmn{pc}^{-2}}\right)^{-1}\rmn{M yr}.\label{tdy}
\end{equation}
 $\overline{\Sigma}_{cl}$ is  the mean surface density of each clump. We also calculate ${\rmn {SFR}}$, assuming  ${\rmn{ SFR}}_{ff}=0.014$ for comparison.

%



\subsection{Relation of Star Formation Rate Surface Density and Gas Surface Density} 
\begin{figure}
	\includegraphics[width=84mm]{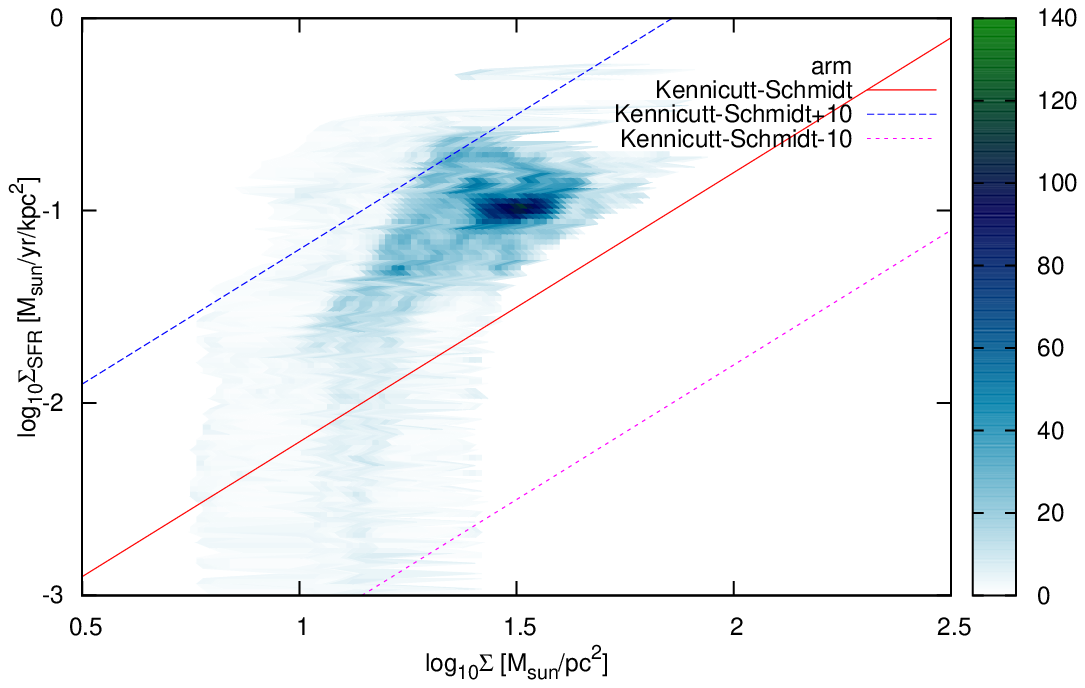}
	\includegraphics[width=84mm]{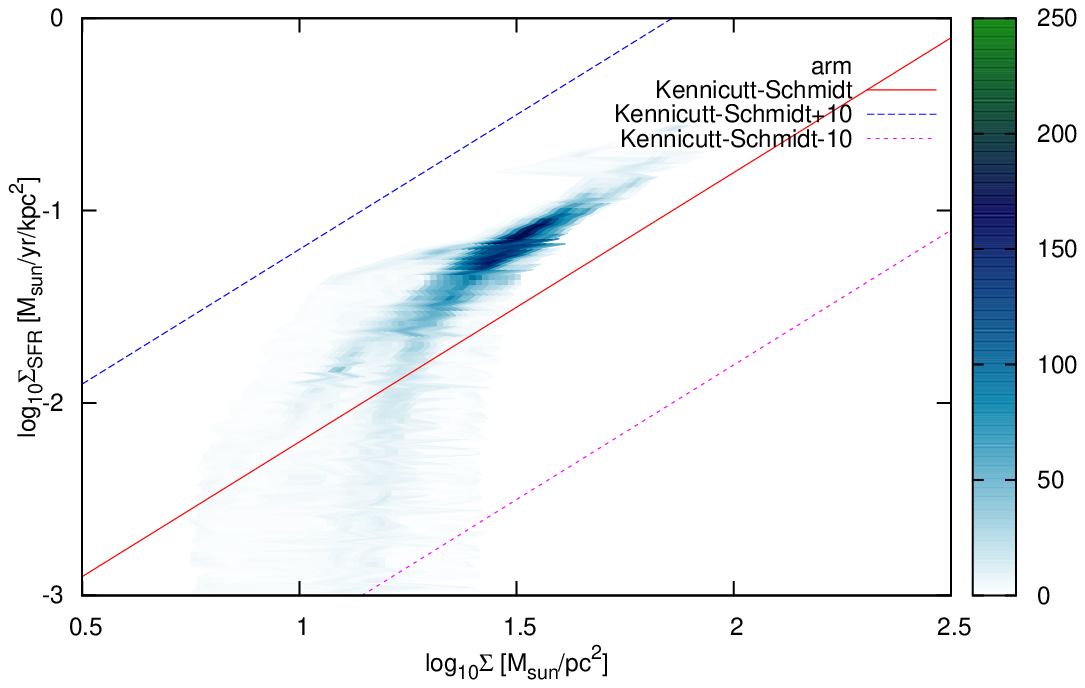}
	 \caption{
	 		The distribution of the SFR surface density  averaged in  the  scale of 500 pc   with  the  clump mass  surface density averaged in the same scale 
			 at $t = $ 120 Myr in the spiral arm region for  SFR$_{ff}$   of eq. (\ref{SFRff}) (top), and  for SFR$_{ff}=0.014$ (bottom). 
			 The color bar shows      number of  cells of numerical hydrodynamic simulation.  
	 	 	The red solid line shows the Kennicutt - Schmidt relation. 
			The blue dotted line shows 10 times the Kennicutt - Schmidt relation. 
			The magenta dotted line shows 1/10 the Kennicutt - Schmidt relation.}
  \label{f:sfd_sfr_3d_s60n12_120}
\end{figure}

\begin{figure}
	\includegraphics[width=84mm]{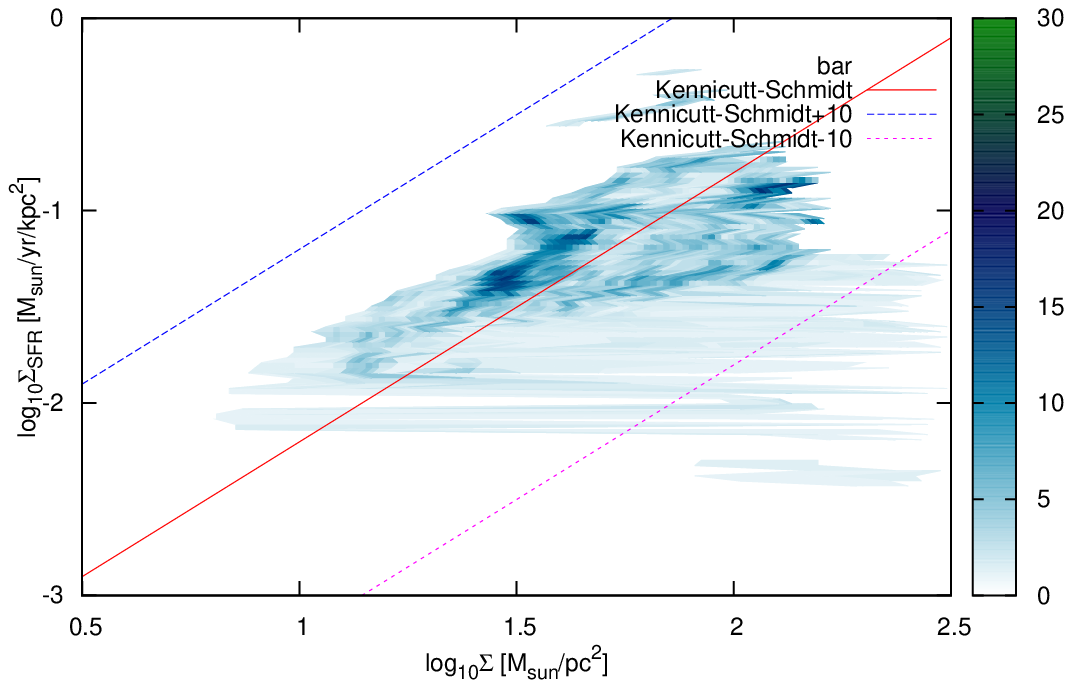}
	\includegraphics[width=84mm]{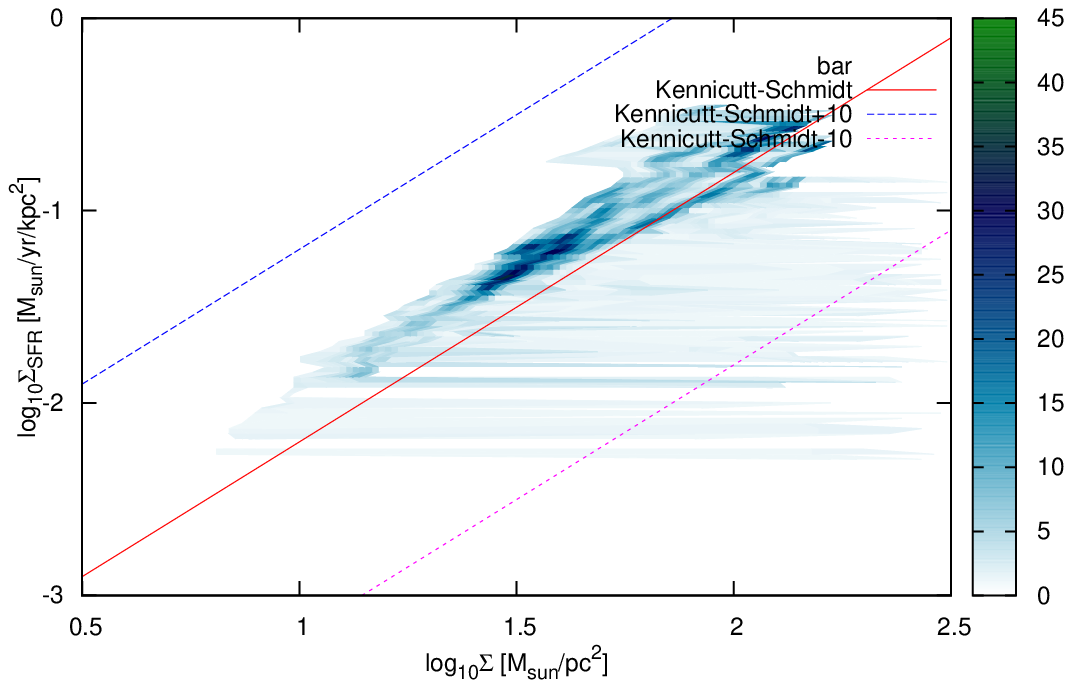}
	 \caption{
	 		The same as in figure \ref{f:sfd_sfr_3d_s60n12_120}, but for the bar regions.}
  \label{f:sfd_sfr_3d_s60n12_120bar}
\end{figure}

We obtain a SFR  from our numerical results by assuming that  the SFR for each clump　is given by equation (\ref{SFR}) with $f_\rmn{MC}=1$, that is,
 the clumps are assumed to be mainly composed  of H$_2$ molecules. 
We show the  SFR  surface density  vs.    surface density of mass of clumps   over     the average scale of 500 pc in Figure \ref{f:sfd_sfr_3d_s60n12_120} (the spiral arm region) and in Figure  \ref {f:sfd_sfr_3d_s60n12_120bar} (the bar region).
 Figure \ref{f:sfd_sfr_3d_s60n12_120} (a) and Figure  \ref{f:sfd_sfr_3d_s60n12_120bar} (a) show  the SFR surface density for  the star formation model with  SFR$_{ff}$ of equation (\ref{SFRff}) as proposed by \citet{km}.  
 These panels show that SFR surface density is lower in the  bar region  than in  the spiral arm region for the same clump mass surface density. This result well corresponds to  the obervational results by \citet{moks}. For the constant SFR$_{ff} $ model  as shown in Figure \ref{f:sfd_sfr_3d_s60n12_120} (b) and Figure  \ref{f:sfd_sfr_3d_s60n12_120bar} (b),  the SFR surface density has smaller scatter  than the result by the model of  \citet{km} in both  regions and difference of the SFR surface density between these regions is small. This result means that the difference of SFR between the bar and spiral arm regions is mainly due to difference of internal motion of clumps, if 
   the star formation model of  \citet{km} can be applied. 
Correlation shown in Figure \ref{f:sfd_sfr_3d_s60n12_120} and Figure \ref{f:sfd_sfr_3d_s60n12_120bar} are well  within  the relation given by    \citet{kn} and  \citet{blwb}.



\subsection{Radial Distribution of Star Formation Rate}
\begin{figure}
  \includegraphics[width=84mm]{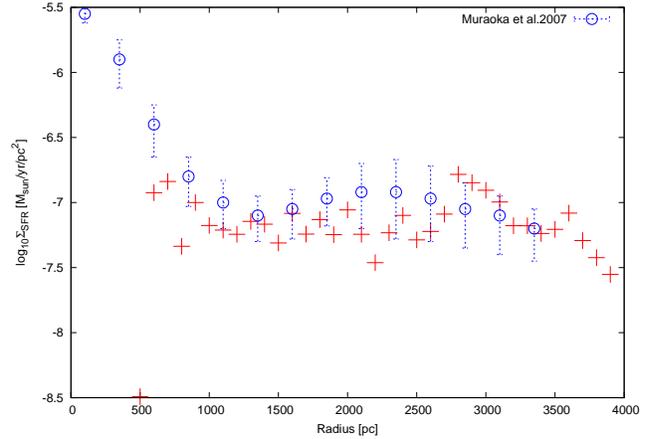}
	 \caption{The radial distribution of star formation rate obtained  by using the Kulmholts-McKee star formation model in the model galaxy . 
	 We show the star formation rate in  $r \ge$ 600 pc.
	 We also plot the star formation rate estimated by \citet{mktk}.
	 	}
  \label{f:rSFR_s60n12_120}
\end{figure}

We show  the  radial distribution of the mean SFR surface density  in Figure \ref{f:rSFR_s60n12_120}. 
In Figure \ref{f:rSFR_s60n12_120}, we exclude the central region of  $r < $ 500 pc, since there is high mass concentration of clumps and the SFR is very high in this region. 
The SFR between $0.5 $ kpc$ ~< r < 2.5$ kpc  is
  $\Sigma_\rmn{SFR} \sim 10^{-7.2}M_{\odot}/pc^2/yr$
 which is  smaller than $\sim 10^{-7}M_{\odot}/pc^2/yr$ in the region of 2.5  kpc$~ < r < $3.5 kpc. 
There are 　small  enhancements of the SFR near    $ r =$2 kpc and  $ r =$3 kpc. The former   is near the outer edge of the bar region,  and  the latter is near the inner ends of gas spirals and  corresponds to small number excess of clumps with $\alpha <1$ as shown in section 3. 
The SFR agrees with    the estimation  of  \citet{mktk} within their uncertainty as shown in Figure \ref {f:rSFR_s60n12_120}. 

\begin{figure}
  \includegraphics[width=84mm]{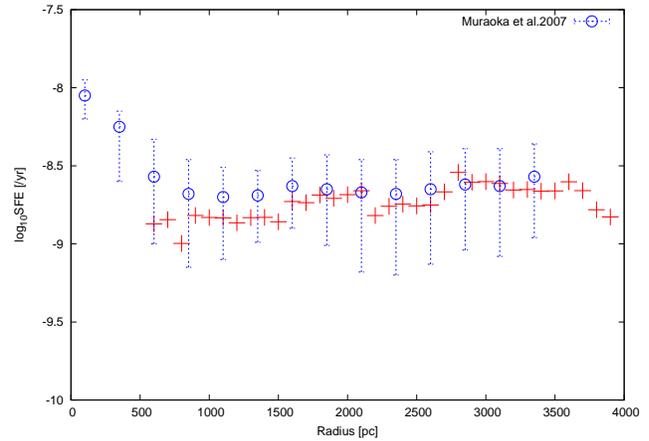}
	 \caption{The radial distribution of star formation efficiency in $r>600$pc..
	 	}
  \label{f:rSFE120}
\end{figure}

We show  the radial distribution of star formation efficiency (SFE), i.e.,  star formation rate per  unit clump mass,   in Figure \ref{f:rSFE120}. 
In this Figure, we also exclude the central region of  $r < $ 500 pc. The SFE  agrees with    the estimation  of  \citet{mktk} within their uncertainty as shown in  Figure \ref {f:rSFE120}. 
The SFE in the bar region is $\sim 60$ \% of  the spiral arm region. 
This trend is observed in barred galaxies ( \citet{mktk}, \citet{moks}). 
Our numerical results show  that many more clumps in the bar region have large virial parameters with $\alpha >1$ than in the spiral arm region.
This is the reason why the SFE is small in the bar region.
 In NGC 4303, SFE  is larger than our numerical results 
with $\sim 10^{-8.5}$ yr$^{-1}$ in  the bar region and $\sim 10^{-8.5}-10^{-7.5}$ yr$^{-1}$ in the spiral arm region.
This is because  SFR  is slightly higher in NGC 4303 than the mean value of the KS relation \citep{moks}.




\section{DISCUSSION and CONCLUSION}
Our numerical results have  shown that most of massive clumps in the bar region have large virial parameters,  $\alpha$ and that mass fraction of bound clumps in the bar region is smaller than in the spiral arm region, although the mean surface gas density is larger in the bar region than in the spiral arm region as shown in section 3. 
  \citet{lowr} shows that molecular gas velocity dispersion  is large  in the bar region  in M83. \citet{sora} shows that molecular gas in the bar has also large velocity dispersion  in a barred galaxy, Maffei 2.
Because of large shear gas flow in the bar region,  most of  clumps formed from gas in the bar region  may have large internal gas motions.
 If cloud collisions in the bar region are more frequent than in the spiral arm region, large internal motion in clouds is also possible to be excited by cloud collisions. 
  These may be the reason why the mass fraction of bound clumps in the bar region is smaller than in the spiral arm  region. 
  It is interesting to study these possibility.
 
The SFR estimation from our numerical results   agrees with the K-S relation as shown in section 4.
In this analysis, we use the star formation  model of  \citet{km} in which they assume that  gravitationally unstable parts of the molecular clouds with   small turbulent velocities  form stars.
In their model, they assume   the scaling relation between turbulent  velocities and   turbulent sizes as in  \citet{l81}.
In their model,  SF rate is given by    the  gravitational unstable part mass  divided by its free fall time. 
 In our analysis, turbulent velocities  of   clump sizes  are assumed to be  the velocity dispersions of internal motions of clumps in our numerical results,　 and equations (\ref{SFR}) and  (\ref{tdy}) are used. 
 We have shown that the SFR surface density is smaller in the bar region than in the spiral arm region for the same gas surface density using the star formation model of \citet{km}. We have also shown that the difference between the bar region and the spiral region is small for the constant SFR$_{\rmn{ff}}$ model. From these comparison, the difference of SFR surface density  can be explained by   the difference of internal motions of clumps between these region.

We have examined the radial distributions of SFR and  SFE, since our estimation of SFR  agrees with the observed Kennicutt-Schmidt relation.
Both radial distributions  are within the error bars of  observation of M83    \citep{mktk} in $r>$ 500pc, with the SFR and SFE being smaller in the bar region than in the spiral arm region.
 The decrease of both  SFR and  SFE in the bar region  is  similar to the recent observations of barred galaxies  \citep{moks}. 

Our numerical results show that the turbulent star formation model 
can  explain the property of star formation in  barred galaxies.
It is interesting  to study observationally the difference in cloud properties in the bar and in the spiral arm regions and the relation between    cloud property and  SF activity   in various  environments in  galaxies using   radio telescopes with  high spatial resolution, e.g. ALMA,  that will  make it possible to  resolve molecular clouds in extragalaxies. 
 

We discuss   limitations of our  numerical simulation. We assume the constant FUV background heating as described in section 2.  Our FUV heating rate that corresponds to our Galaxy value has small effect on cold dense gas, as shown in section 3. The assumption of the constant FUV background heating is too simple.
\citet{th} shows the radial profile  of the surface brightnesses in  the FUV and NUV bands in M83. The FUV surface brightness   is roughly constant from $r=$ 600 pc to $r=$ 2 kpc, has a peak  near $r = $2.6 kpc ( $\sim$2 arcminits ),  and decreases with radius in $r >2.6 $ kpc. This profile  indicates that FUV background heating is stronger in the bar region ($r< 2.6$ kpc) than in the arm region.  If FUV heating is effective to destroy molecular clouds, SFR in the bar  region will be reduced. In this case,  the decrease of   SFR in the bar region  in M83 can be explained  even without the turbulent SF model given by using equation (\ref{SFRff}).  However, It is not obvious that the observed FUV brightness in M83 is strong enough to suppress
SFR in the bar region.
We do not consider   radial-dependent  FUV background heating  in this paper. 
This will be considered in further work. 
 In our simulation, we do not consider feedback  from star formation. Feedback by energy released from supernovae and stellar winds can destroy nearby molecular clouds. 
 Strong UV photons from newly formed stars will ionize nearby  clouds and the ionization will suppress SF in them. 
 Strong stellar winds and supernova  remnants  will  compress molecular  clouds. 
 By the compression, star formation can be triggered.  
 Feedback process is very complicated.
  If  suppression of star formation by feedback is large, difference of SF property between   the bar region and the spiral arm region can be reduced. 
   It is very interesting to study  how feedback affects molecular cloud property  and SF process.
In order  to consider feedback process, 3D numerical simulations with high resolution are essential, since the destruction process  of molecular clouds and expansion of gas by heating of newly formed stars and supernovae are  three dimensional phenomena ( e.g. \citet{tb06}). We will study feedback effects in barred galaxies by three dimensional simulations  in  our forth coming papers.

\section{ACKNOWLEGMENTS}
We  acknowledge Elizabeth Tasker for useful comments that helped  to improve this paper. 
We acknowledge the anonymous referee who gives nice comments to improve this paper. We also acknowledge useful discussions with Masayuki Fujimoto, Tetsuhiro Minamidani,  Naomasa Nakai, Junya Ito, Junichiro Enomoto and  Takashi Ito.   Numerical computations were carried out on Cray XT4 at Center for Computational Astrophysics, CfCA, of National Astronomical Observatory of Japan.  This work was supported by a Grant-in-Aid for Specially Promoted Research 20001003 (AH).



\label{lastpage}

\end{document}